==================
\documentstyle[12pt]{article}

\let\bet=\beta
\let\ga=\gamma

\let\<=\langle
\let\>=\rangle

\newcommand{\s}[1]{#1 \!\!/}
\def\comment#1{ \hbox{Comment suppressed here.} }
\def\tr{\mbox{Tr}\,}

\makeatletter
\def\eqnarray{\stepcounter{equation}%
              \let\@currentlabel=\theequation
              \global\@eqnswtrue
              \global\@eqcnt\z@
              \tabskip\@centering
              \let\\=\@eqncr
              $$%
 \halign to \displaywidth\bgroup
    \eqnumphantom\@eqnsel\hskip\@centering
    $\displaystyle \tabskip\z@ {##}$%
    &\global\@eqcnt\@ne \hskip 2\arraycolsep
         \hfil$\displaystyle{##}$\hfil
    &\global\@eqcnt\tw@ \hskip 2\arraycolsep
         $\displaystyle\tabskip\z@{##}$\hfil
         \tabskip\@centering
    &{##}\tabskip\z@\cr}
\def\eqnumphantom{\phantom{(\theequation)}}

\newcommand{\be}{\begin{equation}}
\newcommand{\ee}{\end{equation}}
\newcommand{\ba}{\begin{eqnarray}}
\newcommand{\non}{\nonumber \\}
\newcommand{\ea}{\end{eqnarray}}
\newcommand{\baa}{\begin{eqnarray*}}
\newcommand{\eaa}{\end{eqnarray*}}
\newcommand{\barr}{\begin{array}}
\newcommand{\earr}{\end{array}}
\newcommand{\bb}{\begin{thebibliography}}
\newcommand{\eb}{\end{thebibliography}}
\newcommand{\ci}[1]{\cite{#1}}
\newcommand{\bi}[1]{\bibitem{#1}}
\newcommand{\lab}[1]{\label{#1}}
\newcommand{\re}[1]{(\ref{#1})}

\newcommand{\PRL}[1]{~Phys.Rev.Lett.~{\bf #1}}
\newcommand{\PR}[1]{~Phys.Rev.~{\bf #1}}
\newcommand{\PL}[1]{~Phys.Lett~{\bf #1}}
\newcommand{\NP}[1]{~Nucl.Phys.~{\bf #1}}

\newcommand{\RMP}[1]{~Rev.Mod.Phys.~{\bf #1}}

\def\ltap{\raisebox{-.55ex}{\rlap{$\sim$}} \raisebox{.4ex}{$<$}}

\def\lsim{\mathrel{\ltap}}

\newcommand{\zn}{$Z_N$~}
\newcommand{\sun}{$SU(N)$~}
\newcommand{\ao}{$A_0$~}

\begin{document}

\thispagestyle{empty}
\setcounter{page}{0}

\begin{flushright}
{\large UBCTP 92-14\\
May 1992}
\end{flushright}
\vspace{1cm}

\begin{center}
\boldmath
{\Large \bf ${Z_N}$ Phases in Hot Gauge Theories}
\footnote{This work is supported in part by the Natural Sciences and
Engineering Research Council of Canada} \\
\unboldmath

\vspace{1.0cm}
{\large Wei Chen},
{\large Mikhail I. Dobroliubov}\footnote{On leave from Institute for Nuclear
 Research, Moscow, Russia},
{\large and Gordon W. Semenoff}\\
\vspace*{0.6 cm}

{\large \it Department of Physics, University of British Columbia\\
Vancouver, B.C., Canada V6T 1Z1}\\
\vspace{0.2in}
\vspace{0.3in}

{\bf Abstract} \\
\end{center}
\vspace*{-5mm}
\noindent

We argue that the \zn phases of hot gauge theories cannot be realized as
a real system with an Hermitean density matrix.

\newpage

It is well known that the equilibrium properties of a system at finite
temperature
can be obtained by the investigation of its evolution in the imaginary time.
The
partition function is given as the functional integral in the Euclidean space
\ci{fin-temp}
\be
Z=\int DA_\mu D\psi D\bar\psi \exp\left\{-\int^\bet_0 d\tau \int d{\bf x}
\left({1\over4}F^2+\bar\psi
D\!\!\!/ \psi\right) \right\}~~~,
\lab{Z}
\ee
with (anti-)periodic boundary conditions for the integration variables.

In hot \sun theory, the effective potential for $A_0$ (or, more correctly,
the expectation value of the Polyakov loop operator)
has been computed \ci{Nathan}
and shown to have local minima for configurations for which the Polyakov loop
operator
$$
L({\bf x})= \tr P\exp\left\{ \int d\tau A_0(\tau,{\bf x})\right\}
$$
lies in the \zn center of \sun , i.e.
$$
<L({\bf x})>\sim \exp \{2\pi i{q\over N} \}~~,~q=0,\ldots,N-1~~~.
$$
They correspond to a uniform  $A_0$ with values
\be
A_0 = \frac{2\pi T}{Ng} \, {1\over 2} \left(
\barr{ccccc}
1 &   &   &   & \\
  & 1 &   &   & \\
  &   & . &   & \\
  &   &   & 1 &  \\
  &   &   &   & -(N-1)
\earr \right) ~~~.
\lab{A0}
\ee
The global minimum occurs at $q=0$, and, depending on the number of flavors of
fermions, there can be local minima at other values of $q$.

One might be tempted to assume that $<\! L\! >$ is an order parameter that
distinguishes thermodynamic phases of the theory.  In that case, although the
global minimum of the effective potential is at $q=0$,
the local minima with $q \neq 0$ would correspond to metastable phases of the
the theory.
A macroscopic bubble with any particular $q$ should be described by a thermal
state or density matrix.
This would assume that it is possible to impose certain constraints in the {\em
real} space on the fields $A_\mu$ entering \re{Z}, so that the Euclidean
description of this constrained system will correspond to only one particular
minimum from \re{A0}. These systems are called \zn phases of the thermal gauge
theory.

Recently \zn phases have been the object of numerous investigations mostly
concerning the evolution of bubbles and their role in cosmology \ci{zn-stuff}.
Later it was realized \ci{prev} that \zn phases at $q=N/2$ (possible for even
$N$) possess bizzare thermodynamic properties, namely for large enough $q$
their fermionic subsystem has {\em positive} free energy, {\em negative}
entropy and {\em negative} specific heat. It has been suggested that these
properties indicate a general instability of \zn phases, or rather that they do
not exist at all as thermodynamics phases in that it is impossible to have a
macroscopic quantity of \zn matter.  Though we agree with the final conclusion,
we find these arguments not fully convincing.

First of all, there are cases with small enough ratio $q/N\lsim 1/4$, (cf.
\ci{prev}) where the fermionic contributions to all thermodynamical quantities
will be normal.  Such \zn phases would not be excluded by the reasoning of
\ci{prev}.

Besides that, the positivity of the contribution to the total free energy of a
subsystem does not necesseraly lead to an instability of the whole system,
because other subsystems can stabilize it. In many cases the contribution of
gluons makes the total free energy of a \zn phase negative (for example, in the
case of QCD the free energy of \zn phases is negative if the number of light
quark flavors is less than 4 \ci{prev}).  To obtain a definite answer to the
question of whether a system is stable or not, it is necessary to analyze the
spectra of its small fluctuations.  We have investigated the properties of the
gluonic and fermionic excitations and found that they don't provide the
tachyonic modes for instability:  the thermal masses and the damping rates of
gauge bosons and fermions in a partucular \zn phase are real and positive.  The
computations are straightforward.  It was shown in \ci{prev} that the $A_0$
condensate can be taken into account by introducing an imagin!
ary chemical potential for the fer
mions, which in turn is equivalent to using Matsubara frequencies $\pi
T(2n+1+2q/N)$.  The gluons are not modified by the condensate.

The thermal masses to leading order in $g^2T^2$ are
\ba
m^2_g = {1\over 3} g^2 T^2 \left[ N + {1\over 2} N_f \left( 1-\left(
\frac{2q}{N} \right)^2_{{\rm mod} 1} \right) \right]~~,
\non
m^2_f = {1\over 8} g^2 T^2 \left[ 1- \left( \frac{2q}{N} \right)^2_{{\rm mod}
1} \right]~~.~~~~~~
\lab{masses}
\ea
(The expression for the gluon thermal mass was first obtained by N.Weiss in
\ci{Nathan}.) Here $C_f=\frac{N^2-1}{2N}$ is the Casimir operator for the
fundamental representation.

It is easy to see that the damping rates of the fermion and gluon excitations,
when their spatial momentum is of order of $T$ (as it is for the most
excitations in the thermal bath), are normal. It was argued \ci{Burgess} that
in the leading $g^2\ln{(1/g)}$ order the damping rates are determined by the
infrared {\em static} behaviour of the corresponding graphs. Obviously, except
for the extremal case $q=N/2$, the fermion propgators in all \zn phases are
infrared finite, as for any $n$ the fermion Matsubara frequency can never be
zero, $p_0=T\pi (2n+1+2 q/N)\neq 0$ for any $n$. Therefore, only the gluon
contributes to the damping rates in this order, similarly in all \zn phases. It
is straightforward to check, that this holds and in the case $q=N/2$, as the
infrared singularity in the fermion propagator is cancelled by $p_0$ from the
fermion loop in the case of gluon damping rate, and by $\s{p}$ in the numerator
for the fermionic one (taking ghosts as an example, one can see!
 that the Bose distribution is nes
secarily but not sufficient to make a particle contribute to the damping rate
at this order).

For the extreme case $q=N/2$ (possible when $N$ is even) one can easily
calculate as well the damping rates at zero spatial momentum. Indeed, as can be
seen from \re{masses}, in this case the fermions do not acquire the thermal
mass in $g^2T^2$ order, therefore there is no need to resum their thermal
loops, which must be resummed, as is the case for other $q$  \ci{Pisarski}. We
obtained
\ba
\ga_g ({\bf p}=0)= \frac{1}{24\pi} (aN+2N_f) g^2 T~~,
\non
\ga_f ({\bf p}=0)=\frac{1}{6\pi} C_f g^2 T~~,~~~~~
\lab{d-r}
\ea
where $a=6.6$ multiplies the gluonic contribution obtained by \ci{Pisarski}.

Thus we see that at the microscopic level the properties of the basic gluonic
and fermionic excitations are qualitatevly the same in all \zn phases and do
not provide modes for instability of the system.
On the other hand, one would expect that a system with negative specific heat
would be unstable to emission of radiation, even if, like the fermions in a \zn
phase, it is a weakly interacting subsystem of a larger system.  A resolution
of this seeming paradox is that  the fermions cannot be treated as an
independent subsystem. This could be done in a weakly interacting plasma, for
example, where the condensate is absent.  However, in the present system it is
the very existence of the condensate, together with the assumption of weak
coupling, which constrains the fermions to have unusual properties\ci{prev}.
There is no natural way to separtate the fermions and gluons in such a
scenario.

However, aside from stability to fluctuations we must consider the question of
whether \zn matter can exist in a macroscopic region.  To do so, it should be
governed by some density matrix or thermal state with sensible properties.  An
indication that the state is metastable would be an imaginary part in the free
energy, which here we would expect to appear when the contributions of
tunneling to the stable q=0 phase is taken into account.  However, at the
perturbative level, one would expect the total free energy as well as other
thermodynamics quantities to be real and display normal thermodynamics.

One such quantity is the particle number density of a particuliar fermion
flavour in a \zn phase. It proves to be {\em complex}, except for $q=0,N/2$ :
\ba
<\Psi^{\dagger}\Psi> = \int \frac{d {\bf p}}{(2\pi)^3} n(p)~~,~~~~~~~~~~~~~~~~
\non
n(p)= 4N\left( \frac{1}{e^{\bet(p-\mu)}e^{2\pi i {q\over N}} +1} -
\frac{1}{e^{\bet(p+\mu)}e^{-2\pi i {q\over N}} +1} \right) ~~~,
\lab{n}
\ea
where for generality we took the real chemical potential $\mu$ to be non-zero
\ci{note}. So, we found that the expectation value of the Hermitean operator
$\Psi^{\dagger}\Psi$ is complex for any $\mu$ and $q\neq 0,N/2$.  Thus, even
for the purpose of analyzing fluctuations, the \zn phase cannot be described by
an Hermitean density matrix or a state in a separable Hilbert space. This
indicates that the state (i.e. a \zn phase) cannot be prepared as a real
macroscopic statistical system, or, equivalently, one cannot impose such gauge
invariant contraints on the fields in the functional integral \re{Z}, which
would pick up only one particuliar minimum from the row  \re{A0}. If the only
constraint is the gauge invariance then one finds all $N$ mimima at the same
time \ci{Nathan}, contributions of which should be sumed up  with the weights
$\exp (-F_q)$, where $F_q$ is the free energy of the $q$`s phase.

There are two possible loopholes in these arguments which could save \zn
phases:
\begin{itemize}
\item The cases with  $q=0,N/2$ have not been excluded, although we do not see
any reason why these values of $q$ can be different from others;
\item If by some reasons (like tunneling, or something else) the true ground
state of the metastable states is a superposition  of the two \zn phases with
$q$ and $(N-q)$, then they will have appropriate thermodynamical properties
(cf. (\ref{masses},\ref{d-r},\ref{n}) and the disscusion in the beginning of
the paper about the macroscpical properties of the fermionic subsystem);
\end{itemize}
Otherwise the possibility of existing domains in the real space, the Euclidean
description of which would correspond to non-zero \ao condensate and \zn
phases, is unambigously excluded.

We are grateful to I.Kogan for fruitful discussions and to N.Weiss, who pointed
out to us on the possibility that \zn phases cannot exist as real statistical
systems.

\bb{99}

\bi{fin-temp} T.Matsubara, Prog.Theor.Phys., {\bf 14}, 351 (1955)

\bi{Nathan} N.Weiss, \PR{D24}, 475 (1981), \PR{D25}, 2667 (1982);\\
D.J.Gross, R.D.Pisarski, and L.G.Yaffe, \RMP{53}, 43 (1981)

\bi{zn-stuff}T.Bhattacharaya, A.Gocksch, C.Korthals Altes, and R.D.Pisarski,
\PRL{66}, 998 (1991);\\
J.Ignatius, K.Kajantie, and K.Rummukainen, \PRL{68}, 737 (1992);\\
V.Dixit and M.C.Ogilvie, \PL{B269}, 353 (1991)

\bi{prev} V.M.Belyaev, I.I.Kogan, G.Semenoff, and N.Weiss, \PL{??}

\bi{Burgess} C.P.Burgess and A.L.Marini, \PR{D45}, R17 (1992);\\
A.Rebhan , CERN preprint CERN-TH.6375/92 (1992)

\bi{Pisarski} E.Braaten and R.D.Pisarski, \PRL{64} 1338 (1990), \NP{B337}, 569
(1990)

\bi{note} It is interesting to note, that integrated over $\mu$, the number
density gives the free energy of the system,
\[
F=\int \frac{d {\bf p}}{(2\pi)^3} \left[ \log{(1+e^{\bet(p-\mu)}e^{2\pi i
{q\over N}})} + \log{(1+e^{\bet(p+\mu)}e^{-2\pi i {q\over N}})} \right]~~,
\]
which is real only in the three cases: either when one sums  $q$ and $(N-q)$
(or over all possible $q$, as is the case for the usual vacuum), or when there
is no real chemical potential, $\mu =0$ (as was considered in
\ci{Nathan,prev}), or, for any $\mu$, when $q=0,N/2$.
\eb
\end{document}